\documentclass[useAMS,usenatbib,usegraphicx]{mn2e}  
\usepackage[usenames]{color}

\newcommand{\be}{\begin{equation}}
\newcommand{\ee}{\end{equation}}
\newcommand{\bea}{\begin{eqnarray}}
\newcommand{\eea}{\end{eqnarray}}

\title[EOSL in MACS0416] 
{The Orthogonally Aligned Dark Halo of an Edge-on Lensing Galaxy in the Hubble Frontier Fields: A Challenge
for Modified Gravity.}
\author[J.M. Diego]  
  {Jose M. Diego\footnote{jdiego@ifca.unican.es}$^1$, Tom Broadhurst$^{2,3}$, Narciso Benitez$^4$, Jeremy Lim$^5$, Daniel Lam$^5$ \\ 
$^{1}$IFCA, Instituto de F\'isica de Cantabria (UC-CSIC), Av. de Los Castros s/n, 39005 Santander, Spain\\
$^{2}$Fisika Teorikoa, Zientzia eta Teknologia Fakultatea, Euskal Herriko 
Unibertsitatea UPV/EHU\\ 
$^{3}$IKERBASQUE, Basque Foundation for Science, Alameda Urquijo, 36-5 48008 Bilbao, Spain\\
$^{4}$Instituto de Astrof\'isica de Andaluc\'ia (CSIC), Apdo. 3044, 18008 Granada, Spain\\
$^5$ Department of Physics, The University of Hong Kong, Pokfulam Road, Hong Kong 
}

\date{Draft version \today}  
  
\pagerange{\pageref{firstpage}--\pageref{lastpage}}  
  
\begin{document}  
\maketitle  
 
\label{firstpage}  
\begin{abstract}  
We examine a well resolved 8\arcsec\   lensed image that is symmetrically bent in the middle by an edge-on lenticular galaxy, in the Hubble Frontier Field (HFF) data  of MACSJ0416.1-20403. The lengthy image is generated primarily by the large tangential shear of the cluster with a local secondary deflection by the member galaxy out to a limiting radius of $\simeq 18$ kpc. The lensing lenticular galaxy is also well resolved and evidently lies nearly edge-on in projection. This fortuitous combination of a long arc intersecting an edge on galaxy provides us with an opportunity to place relatively strong constraints on the lensing effect of this galaxy. We can model the stellar lensing contribution using the observed pixels belonging to the galaxy, in 2D, and we add to this a standard parameterised dark halo component. Irrespective of the detailed choice of parameters we obtain a combined total mass of $\approx 3\times 10^{11} M_{\odot}$. Depending on the dark halo parameters, the stellar contribution to this is limited to the range $5-15\times 10^{10}M_{\odot}$, or 20-50\% of the total mass, in good agreement with the independent stellar mass computed from the photometry of $5\times 10^{10}M_{\odot}$ for a Chabrier IMF, or $8\times 10^{10}M_{\odot}$ for a Salpeter IMF. The major axis of the DM halo is constrained to be nearly orthogonal to the plane of the galaxy, within a range of $\sim 15^{\circ}$, and with an ellipticity $e \approx 0.15$ corresponding to an axis ratio $a/c=0.54$. We show that these conclusions are very weakly dependent  on the model of the cluster, or the additional influence of neighbouring galaxies or the properties of the lensed source. Alternative theories of gravity where the radial dependence is modified to avoid the need for DM are challenged by this finding since generically these must be tied to the baryonic component which here is a stellar disk oriented nearly orthogonally to the lensed image deflection. Other such fortuitously useful  lenses can be examined this way as they become uncovered with more HFF data to help provide a more statistical distribution of galaxy halo properties.
\end{abstract}  
\begin{keywords}  
   galaxies:clusters:general;  galaxies:clusters ; dark matter  
\end{keywords}  
  
\section{Introduction}\label{sect_intro}  

Data from the Hubble Frontier Fields \footnote{http://www.stsci.edu/hst/campaigns/frontier-fields/} 
program (HFF hereafter) is providing exquisite precise observations of lensing distortions 
around galaxy clusters. Clusters observed under this program
routinely contain tens of multiply lensed systems with well defined
colours that help identify counter-images. The abundance of strong
lensing observables allows for an unprecedentedly detailed reconstruction of the dark 
matter (DM) distribution. 
One of these clusters, MACSJ0416.1-2403, at $z=0.4$, was recently studied by us using the Hubble 
data. A precise model for the distribution of DM was obtained based on the gravitational 
lensing effect (\cite{Diego2014}). 
Among other results, we found yet another example of a cluster halo without a pronounced cusp at its centre 
while numerical simulations predict that CDM form haloes with the density monotonously increasing towards the centre.

The tremendous success of the cold dark matter model is being challenged by these types of inconsistencies between  
observations and predictions. 
Alternative models have been proposed invoking changes in the laws of gravity \cite{Milgrom1983,Sanders2002,Clifton2012,Khoury2014} that 
aim at explaining the observations without invoking the existence of DM. 
Some of the most popular alternative theories to CDM adopt a modified version of the laws of gravity and assume that 
there is no DM and the source of the gravitational force is in the baryonic matter itself. 
These alternative models had relative success in explaining some observations like rotation curves of galaxies but 
they have also been seriously challenged by observations of galaxy clusters where the source of the gravitational 
potential does not seem to always follow the bulk of the baryon component. 
One of these clusters is the {\it bullet} cluster where two 
clusters have crossed each other separating the plasma from the galaxies. As the plasma contains most of the baryonic 
mass in the cluster, these alternative models clearly predict the minimum of the gravitational potential to be at 
the position of the plasma, which contains most of the baryonic mass in the cluster. 
Instead, detailed analyses of the gravitational lensing effect around this cluster reveal 
that the minimum is located where the CDM model predicts, around the galaxies and their associated, and 
{\it invisible}, DM haloes. Other clusters show similar patterns, suggesting that the galaxies 
must have a large portion of DM around them. 
What these observations can not tell in detail is the exact distribution of DM. Hence, it is still (in principle) 
possible that the source of the gravitational potential is associated with the galaxies rather than the diffuse gas, or DM halo.  
Unfortunately, the analysis of the gravitational lensing effect suffers important degeneracies. Results derived from gravitational lensing effect 
are very accurate in the prediction of the mass contained within certain radius, but in most cases they can not provide 
accurate additional information about how the mass is distributed within that radius. 

As a result, some room still exists for alternative theories where the baryons (i.e the galaxies) 
are still the main source of the gravitational potential. Intriguingly, the success of lensing reconstruction 
methods based on the simple assumption that light traces matter suggests the existence of a strong link between the 
baryons and the dark matter while in theory, at the scales of the galaxies, dark matter haloes are expected to 
adopt triaxial shapes which do not necessarily need to be aligned with the galaxies they host 
(as suggested by numerical N-body simulations).  The test proposed on this work and based on edge-on secondary lenses (EOSL),  
offer a unique opportunity to test cases where the baryons adopt an extreme distribution. 
If the baryons are the source of the gravitational field, whether gravity obeys 
a standard $1/r^2$ law or not, the extreme geometry should be reflected in the lensed background galaxy. If on the other 
hand, baryons are not the main source of the gravitational field (as expected for the standard LCDM model), the EOSL galaxy 
may reveal the need for a halo of dark matter surrounding the galaxy which is morphologically distinct from the peculiar 
geometry of the EOSL and that would directly challenge theories of modified gravity. 
EOSL galaxies take advantage of the lensing power of the galaxy cluster to stretch a background galaxy. The magnifying power of the host cluster 
stretches the background galaxy to produce nearly straight and featureless arcs. These elongated arcs are ideal background sources when re-lensed 
by a secondary lens. We note that the lensing distortion does not distinguish between the deflection from the host cluster and the secondary 
lens but mathematically we can separate the two effects. When the two lenses (cluster and EOSL) are in the same lens plane, 
the deflection angle, which is an integrated effect, can be viewed as a linear process where 
a background galaxy is first lensed by the host cluster and later lensed again by the secondary lens. 
In selected regions of the cluster where straight arcs are more likely to appear, the dominant deflection field from the galaxy cluster is able to transform an 
intrinsically small background galaxy (with unknown intrinsic shape) into an elongated arc with a shape that can be well approximated by a straight elongated 
line. If this arc is elongated enough, it can provide a wide range of angular distances over which one can test the deflection field from the EOSL. 

\begin{figure}  
{\includegraphics[width=8.5cm]{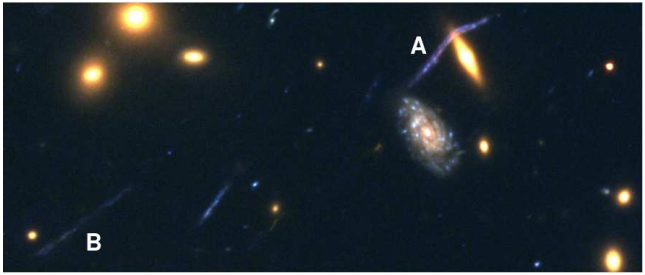}}
   \caption{The background galaxy lensed by the EOSL is marked with A. The EOSL galaxy is shown saturated here but it exhibits 
            a thin disc and bright nucleous. 
            Another very elongated straight arc nearby is marked with a B. The orientation of arc B agrees well with 
            the expected orientation from the lens model.   
            The field of view is $31.4\times13.2$ arcseconds$^2$.}
   \label{fig_Data}  
\end{figure}

An example of this effect is shown in figure \ref{fig_Data}.  Two prominent, and very elongated arcs, marked A and B, can be seen in this 
image. The arc denoted as B is found in the south-west sector of the cluster MACSJ0416.1-2403 and beyond the Einstein radius. In the case of arc B, 
the distortion is mostly due to the host cluster with no nearby secondary lens affecting the arc. 
Using the lens model derived by \cite{Diego2014}, we are able to predict the estimated orientation produced by the lens model at this position and 
we find that the predicted and observed orientations agree to within 3 degrees. The straight nature of the arc B is 
an indication of the smoothness of the deflection field of the cluster in this region of the lens but also of its orientation. 
While no nearby galaxies are found near arc B that could distort this smoothness, the contrary occurs in arc A. In this case, 
the straight arc gets significantly distorted by the presence of a galaxy that adds a distortion to the smooth deflection 
field from the cluster. The difference between arcs A and B summarizes in a graphical way the power of the test proposed 
in this work. The cluster contributes to the effect by producing a very elongated nearly-straight arc with an orientation 
that can be constrained by the derived deflection field from the cluster (like in arc B). A galaxy member can introduce a correction to 
this effect and distorts the straight arc (like in arc A). This distortion depends, mostly, on the mass distribution of 
the EOSL. The power of this test resides in the combination of the two effects. The cluster produces an image with a 
known shape (a straight line with a well constrained orientation) reducing one of the uncertainties in lensing 
reconstruction (the unknown intrinsic shape or orientation of the background galaxy). The magnifying power of the cluster 
produces also large elongated arcs. Long arcs are useful to constrain the shape of the mass distribution of the 
secondary lens over a wider range of distances. As opposed to Einstein rings that mostly constrain the mass enclosed in the ring 
(that is in the radial direction), the arc A aligns in a perpendicular direction to the lens effectively proving the two 
directions, radial and tangential, and thus expossing the real geometry of the lens. 
The EOSL adds a local correction to the deflection field that can be used to constrain the distribution of the mass in the member galaxy. 
If the member galaxy is found to be edge-on (like the member galaxy lensing arc A in figure \ref{fig_Data}), then 
the sensitivity to the mass in the halo versus the mass in the baryons gets maximized as the DM is not expected to 
concentrate in disc-like formations. Collisionless DM behaves very different than baryonic matter and tend to form 
triaxial haloes, not discs. If the best model for the EOSL is found to concentrate most of the mass in the disc region, 
this would have profound implications for the nature of DM since it would require DM to share some of the properties of 
baryons (like self-interaction) in order to explain such a distribution or provide inmense support to theories of modified gravity. 
If on the contrary, the best model for the EOSL requires an extended halo of DM around the galaxy with a very different spatial distribution, this result would 
represent a new challenge for models of modified gravity that require the baryonic matter to be the source of the 
gravitational potential. 

In this paper we apply the EOSL test to the lensed galaxy A shown in figure \ref{fig_Data} and derive constraints 
on the mass distribution of the EOSL galaxy. The EOSL galaxy
is located at RA=04:16:04.094, DEC=-24:05:22.44 (J2000)\footnote{We named the background lensed galaxy the {\it Dragon Kick}}. 
In section \ref{method} we describe the method used in this paper. In section \ref{results} we present the main results 
from our analysis. We discuss possible sources of systematics and alternative models in section \ref{resultsII}   
Through the paper we assume a cosmological model with $\Omega_M=0.27$, $\Lambda=0.73$, 
$H_o= 70$ ${\rm km/s/Mpc}$. For this model, 1 arcscec equals $5.42~kpc$ at the 
distance of the cluster. 

\section{Methodology}\label{method}

\begin{figure}  
{\includegraphics[width=8.5cm]{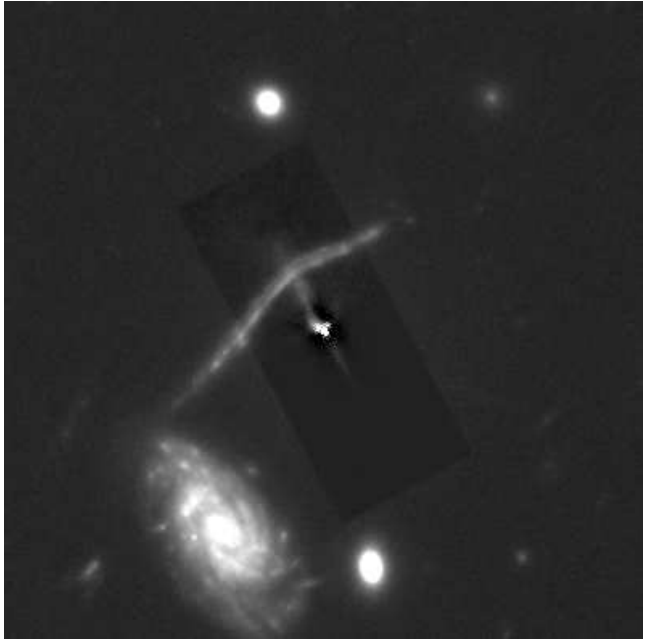}}
   \caption{Background lensed galaxy after partially removing the foreground galaxy (lens). The field of view 
            is 12 arcseconds across.}
   \label{fig_Data_clean}  
\end{figure}

Our base model for the lens is based on two components. 
The first one accounts for the baryonic distribution and the second one accounts 
for a hypothetical DM halo around the galaxy. 
The distribution of the baryonic component is assumed to follow the observed flux of the galaxy. 
As can be seen in figure \ref{fig_Data}, the EOSL has a very symmetric distribution. 
Taking advantage of this symmetry, we used the SW half of the galaxy to subtract 
the NE half under the lensed background galaxy. The result is shown in figure \ref{fig_Data_clean}. The part 
of the galaxy that has been used to clean the other half has been lightly smoothed with a Gaussian before subtraction 
in order to maintain the small scale details in the difference. As a consequence of the smoothing the subtraction is not perfect and a 
small residual is left, specially in the disc plane of the galaxy and its centre. The residual also reveals that the 
EOSL galaxy is not perfectly symmetric with the NE section having a slightly more flux than the SW half. 
The distribution of the DM component is assumed to follow a triaxial distribution. In this work we consider 
prolate models ($a=b$ and $c>a$) although we have checked that our results are insensitive to whether the DM halo 
is prolate or oblate  ($a=b$ and $c<a$). 
The degree or prolateness can be described with a single parameter, the ellipticity. 
The ellipticity, $e$, is defined as; 
\begin{equation}
e=\frac{c-a}{2(a+b+c)}
\label{eq_e}
\end{equation}
with $a=b$, $b<c$ and $c$ is the longest axis of the ellipsoid (in the oblate case $c$ would be the shorter axis). 
According to \cite{Despali2014}, when defined this way, $|e|$ takes values between 0 and 0.3 
with $e=0.1$ being a typical value. See also values from \cite{Jing2002}. Positive values of $e$ correspond to 
prolate models while negative values of $e$ correspond to oblate models.  

In addition to the ellipticity, we allow the DM halo to have a given orientation, $\alpha$, with respect to the normal of the EOSL disc. 
That is, the value  $\alpha=0^{\circ}$ corresponds to the case when the longest axis, $c$, is perpendicular to the plane of the galaxy disc. When $\alpha=\pm 90^{\circ}$ the longest axis 
of the triaxial DM halo is aligned with the galaxy disc. For the mass distribution we adopt a triaxial NFW profile (but with the two axis, $a$ and $b$ taking identical 
values). 
We simulate the DM spheroids by projecting the 3D triaxial NFW profile with the three axis re-scaled by 
the corresponding $a,b,c$ for a given $e$. If $a=b$ and $c=1$, given $e$, $a$ (or $b$) is determined as, 
$a=(1-2e)/(1+2e)$. The values of the NFW concentration parameter and virial radius are fixed to $C=3$ and $R_{200}=15 kpc$ respectively. Other values will be discussed later. 

When including the masses of the EOSL and the surrounding DM halo, our lens model is then described by 4 parameters, $M_{EOSL}$, $M_{DM}$, $e$ and $\alpha$. 
$M_{EOSL}$ refers to the total baryonic mass that has a distribution that follows the observed flux of the EOSL. $M_{DM}$ accounts for the total mass of the triaxial 
halo around the EOSL, $e$ is the ellipticity of the triaxial halo which can take values from $e=0.0$ to $e=0.3$ and $\alpha$ accounts for 
the orientation of the halo with values of $\alpha \sim \pm 90^{\circ}$ corresponding to the particular case where the DM halo and the EOSL are aligned 
(this case mimics the scenario where the DM halo is oblate but with the longest axis aligned with the EOSL disc) and 
$\alpha \sim 0^{\circ}$ for the case when they are perpendicular (like in figure \ref{fig_geom_lens}).   
Values of $e \approx 0$ correspond to DM haloes that are nearly spherical. 

\begin{figure}  
{\includegraphics[width=8.5cm]{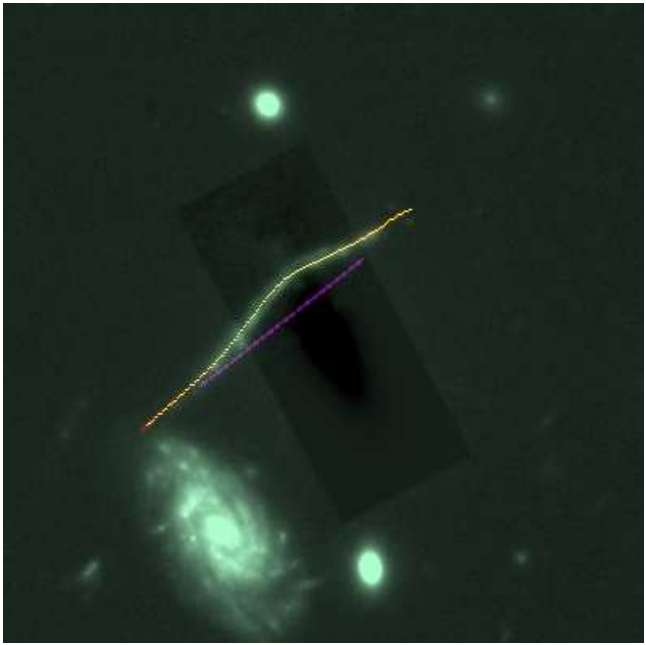}}
   \caption{Lensed galaxy compared with a typical good fit model showing the background source in purple.}
   \label{fig_Data_vs_bestfit}  
\end{figure}  

In addition to modeling the lens, a model is needed for the background source. In this case we take advantage of the 
dominant magnifying power of the cluster. The dominant lensing effect of the cluster near the position of the arc is twofold. 
First, the cluster introduces an orientation in background lensed sources. This can be appreciated 
for instance in arc B in figure \ref{fig_Data}.  Second, the magnifying power of the cluster stretches the background source 
as can be appreciated also in arc B in figure \ref{fig_Data}. The combined effect produces an ideal source to be {\it relensed} 
by a secondary lens that offers a large cross-section for the lensing effect but also a nearly constant inclination of the background 
source. Using the lens model derived in \cite{Diego2014} (in particular, we used 
case II in that paper where the EOSL galaxy was included as part of our model), we can estimate the expected average orientation 
of the background source at the position of the observed arc (after removing the effect of the EOSL from our lens model). 
The model predicts that a background source that is lensed into this position should have an 
inclination of 37 degrees with respect to the horizontal line in figure \ref{fig_Data}. As any other model, the one 
used in this work is not free of error so this estimate should be taken as an approximate value. However, by looking 
at nearby arcs we find that this estimate is close to what is observed in neighbouring arcs. For instance, the nearby 
arc B in figure \ref{fig_Data} has an inclination of 34 degrees with respect to the horizontal line. 
Also, small departures from the predicted inclination are expected due to the intrinsic inclination of the background 
source. However, this should be only a minor correction when compared with the inclination induced by the dominant lensing effect 
of the cluster. To include the uncertainty in the orientation of the background source (after accounting for the 
cluster-induced orientation) and uncertainties in our cluster lens model we explore a range of orientations, $\theta$. This variability 
is also recommended to account for possible errors in the lens model that could be produced by incomplete sky coverage 
of the lensing constraints (more concentrated around the centre of the cluster) or missing substructure in our lens model. 
This second possibility will be considered in more detail in section \ref{resultsII}.

The position of the background source with respect to the centre of the EOSL is also varied since there is a 
degeneracy between the lens mass and the source position that needs to be accounted for. The source model is then described 
by two parameters, the inclination, $\theta$, and the separation (or impact parameter), $\Delta d$, 
with respect to the centre of the EOSL. For convenience we choose  $\Delta d$ not with respect to the centre of the EOSL but with respect to the 
{\it zero point} (see below) that lies in between the centre of the EOSL and the background lensed galaxy. 
The global effect of the cluster is taken into account by modeling the background source as an elongated arc with a 
given length and a given inclination ($\theta$ parameter above). 
Finally, and also determined by the cluster is the {\it zero position} of the lensed arc. 
This position is defined as the position at which we would see the straight 
arc if the EOSL was not in between the observer and the arc. For clarity we show in figure \ref{fig_Data_vs_bestfit} 
the a source model (cyan straight line) before being lensed by the EOSL. The cyan line is how the galaxy should 
have been observed if only the cluster and substructure not included in our lens model were acting as the gravitational lens. 

\begin{figure*}  
{\includegraphics[width=19.5cm]{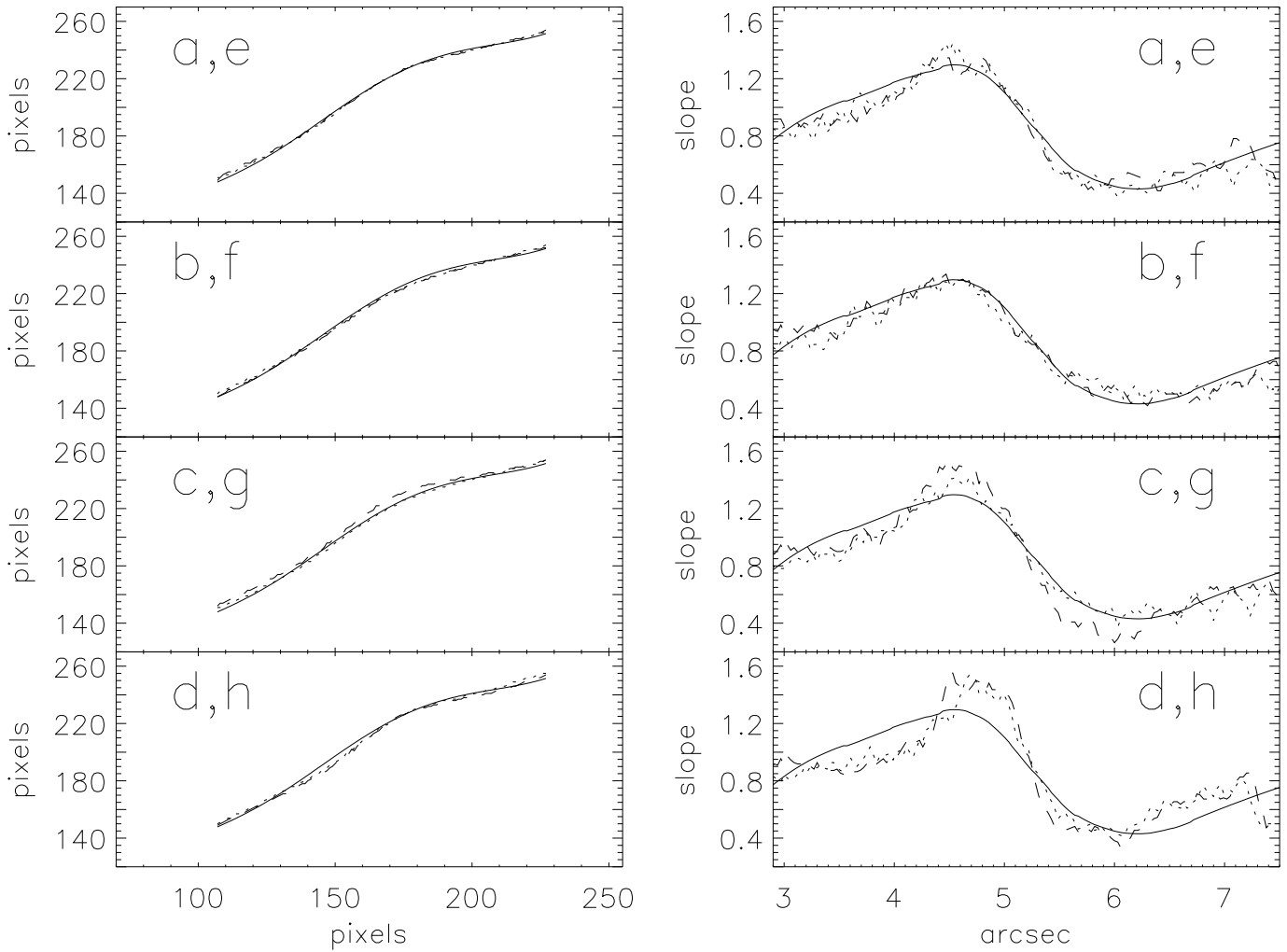}}
   \caption{Best fitting models listed in table 1 compared with data. The dotted lines correspond to the set of models a,b,c,d (obtained after 
            fixing the mass of the spiral galaxy to $M_{spiral}=5\times10^{10}M_{\odot}$). The dashed lines are for the set of models e,f,g,h 
             (obtained after fixing the mass of the spiral galaxy to $M_{spiral}=22\times10^{10}M_{\odot}$). The solid line shows the data used in the fit.}
   \label{fig_Fit_grid}  
\end{figure*}  
 
We vary the 6 parameters described in the previous section, $M_1,M_2,e,\alpha,\theta,\Delta d$, and compare the observed arc 
with the one predicted for each model. We define a standard $Log(L)=-\chi^2$ to select the best models where; 
\begin{equation}
\chi^2 = \chi^2_{slope} + \chi^2_{pixel}
\end{equation}
The terms $\chi^2_{slope}$ and $\chi^2_{pixel}$ are defined as a standard quadratic difference between the data and 
the model where in $\chi^2_{slope}$ the data (model) correspond to the observed (predicted) slope of the arc and 
in $\chi^2_{pixel}$ the data (model) correspond to the observed (predicted) position of the arc. In Figure \ref{fig_Fit_grid} we show the data 
compared with a few selected models.  The errors in both terms are chosen to make 
their contributions to the global $\chi^2$ of comparable importance around its minimum and also to make the reduced $\chi^2$ of order unity. 
We found that adding the slope is a sensitive 
measure since it is able to account for small changes in the predicted curvature that may not be reflected so well when 
using only a $\chi^2$ based on the difference of positions. Also, the slope is less sensitive to substructure not accounted for in our model and that 
is further away from the EOSL. The main effect of the substructure is to add an additional gradient to the deflection field around the observed arc. 
This gradient has a small effect on the curvature near the observed disc of the EOSL galaxy, which is the region where the slope changes most. 

In figure \ref{fig_Fit_grid} we show both the slope and position of the arc used to build the $\chi^2$ compared with the predicted position and 
curvature of a set of models that will be discussed in more detail later. The curvature is more sensitive to the details of the mass distribution near the galaxy disc. 
This measurement is important to discriminate between models that contain the mass concentrated in the disc or spread over a wider region. 

\subsection{The effect of missing substructure}\label{resultsII}   

    \begin{table*}
      \centering
    \begin{minipage}{130mm}                                               
    \caption{Models represented in figure \ref{fig_Fit_grid}. Models a,b,c,d correspond to the case where the mass of the spiral galaxy is fixed to $5\times 10^{10}M_{\odot}$.
             Model a corresponds to the global  best fit. Model b is the best model when the EOSL mass is forced to take the lowest value of our grid of parameters. 
             Model c is the best model when the DM ellipticity is forced to be very small ($e=0.01$). 
             Model d is the best model when the mass of the DM halo is forced to be zero. 
             Models e, f, g, h are the equivalents to a,b,c,d but when the mass in the spiral galaxy is fixed to 
             $22.0\times 10^{10}M_{\odot}$. The last column shows the value of the $\chi^2$ relative to the $\chi^2$ of the best model (a). The quoted errors correspond 
             to half the bin width in our multidimensional grid.}
             
 \label{tab1}
 \begin{tabular}{cccccccc}   
  \hline
   Model  &   $\theta(^{\circ})$   & $\Delta p (pixel)$ &    $e$   & $\alpha(^{\circ})$ &  EOSL($10^{10}M_{\odot}$)   & DM($10^{10}M_{\odot}$) & $\chi^2$ \\
          &    $\pm 1$   &  $\pm 1$ & $\pm 0.025$&$\pm 7.5$&$\pm 0.4$&$\pm 1.0$&     \\
  \hline
    a     &     32       & -2         & 0.15     &     0   &  11.02   & 18.57 &   1.00 \\
    b     &     32       & -2         & 0.1      &     0   &   2.86   & 28.77 &   1.38 \\
    c     &     32       & -2         & 0.01     &     0   &   4.49   & 32.85 &   1.15 \\
    d     &     36       &  4         &   -      &     -   &  18.97   &  0.0  &   2.97 \\
  \hline
    e     &     38       & -2         & 0.15     &     0   &  12.65   & 14.49 &   1.18 \\
    f     &     38       & -6         & 0.1      &   -15   &   2.86   & 32.85 &   1.46 \\
    g     &     38       & -2         & 0.01     &   -60   &  14.28   & 24.69 &   2.01 \\
    h     &     40       & 0          &   -      &    -    &  21.42   &  0.0  &   3.27 \\
  \hline
 \end{tabular}
 \end{minipage}
\end{table*}

From figure \ref{fig_Data}  it is evident that the spiral galaxy to the south east of the EOSL may have a non-negligible effect. 
An estimation of the stellar mass of this spiral galaxy (and the EOSL) can be obtained based on the photometry. 
The stellar mass is derived after assuming an initial mass function (IMF). We adopt a standard Chabrier IMF to estimate the stellar masses of 
the spiral galaxy and find $M_{spiral}=(1.3 \pm 0.3)\times 10^{10} M_{\odot}$. 
We estimate also the stellar mass of the EOSL. We obtain $M_{EOSL}=(5.0 \pm 1.1)\times 10^{10} M_{\odot}$ 
(for the Chabrier IMF). For a Salpeter IMF the mass is slightly larger, $M_{EOSL}=8 \pm 2 10^{10} M_{\odot}$.
These values will be discussed later in section \ref{results}. The mass of the spiral is then expected to be smaller than the mass of the 
EOSL but in principle not negligible. 
In addition, in a recent paper, \cite{Jauzac2014}, the authors present a lens model based on combined 
weak and strong lensing data with the weak lensing data covering a larger field of view that reaches beyond the EOSL. 
They find evidence for a substructure, denoted by S1 in their paper, and located $\approx 140$ kpc south of the EOSL. 

If these substructures have significant masses they may introduce an additional gradient in our lens model. Since both, the spiral galaxy and S1, are located to the south 
of the EOSL, the effect of the gradient would be to reduce the  inclination angle for the source. This can be appreciated easily if we consider the extreme, albeit unrealistic, 
case where the dominant effect is due to the spiral and/or S1 substructure instead of the cluster. In this case the inclination angle would be $\theta \approx 0^{\circ}$ 
since the substructure is immediately below (south) the EOSL and its effect on the background galaxy would be to stretch it horizontally (i.e  $\theta \approx 0^{\circ}$). 
If the spiral and S1 has a non-negligible effect at the position of the EOSL, we should expect the inclination angle to be smaller than the predicted 
 $\theta=37^{\circ}$ (that is, somewhere between 37 and 0 degrees). 
In fact, if we perform our fit ignoring the effect of substructure like the spiral galaxy and the structure S1 in the south, the best fitting models show a strong preference for values 
of $\theta$ that are almost 10 degrees lower than the predicted value of  $\theta=37^{\circ}$. A degeneracy between $\theta$ and the mass of the spiral galaxy and/or S1 is expected. The 
exclusion of the spiral galaxy in our model results in a smaller than expected $\theta$, 
that accounts for the missing structure, 
in good agreement with the hypothesis that the spiral and/or S1 
play a non-negligible role in the determination of the tilt angle. 
Although in our tests we place a halo at the position of the spiral and the structure S1 is neglected explicitly, 
it is important to note that any significant effect from S1 can be mimicked by assigning a larger mass to the spiral galaxy.
The mass of the spiral galaxy should then be considered as an effective mass that takes into account the masses of the spiral and S1 together with their respective distances 
to the lensed arc. The role of the spiral galaxy is tested in our model by including a spherical DM halo at the position of the spiral. We consider two scenarios. 
In the first one we adopt a relatively 
small mass for the spiral of $5\times10^{10} M_{\odot}$. This mass is approximately 4 times the mass of the stellar component estimated with a Chabrier IMF and should be considered as a reasonable lower limit for 
the total mass of the spiral galaxy. In the second scenario, we assign a mass to the spiral of $2.2\times10^{11} M_{\odot}$. 
This mass is particularly high but not unreasonable if we keep in mind that the spiral galaxy accounts also for the effect of the S1 structure.
In both cases we assume a NFW spherical halo with concentration $C=7$ and $R_{200} = 30$ kpc. 
The particular choice of shape, concentration and virial radius has a very small impact on the results, with the total mass being the most important parameter. 
Given their location in relation to the EOSL, both masses, from the spiral and S1 structures, are expected to be degenerate so a high derived 
mass in the spiral may be due to the non-negligible effect of S1. 

\begin{figure}  
{\includegraphics[width=8cm]{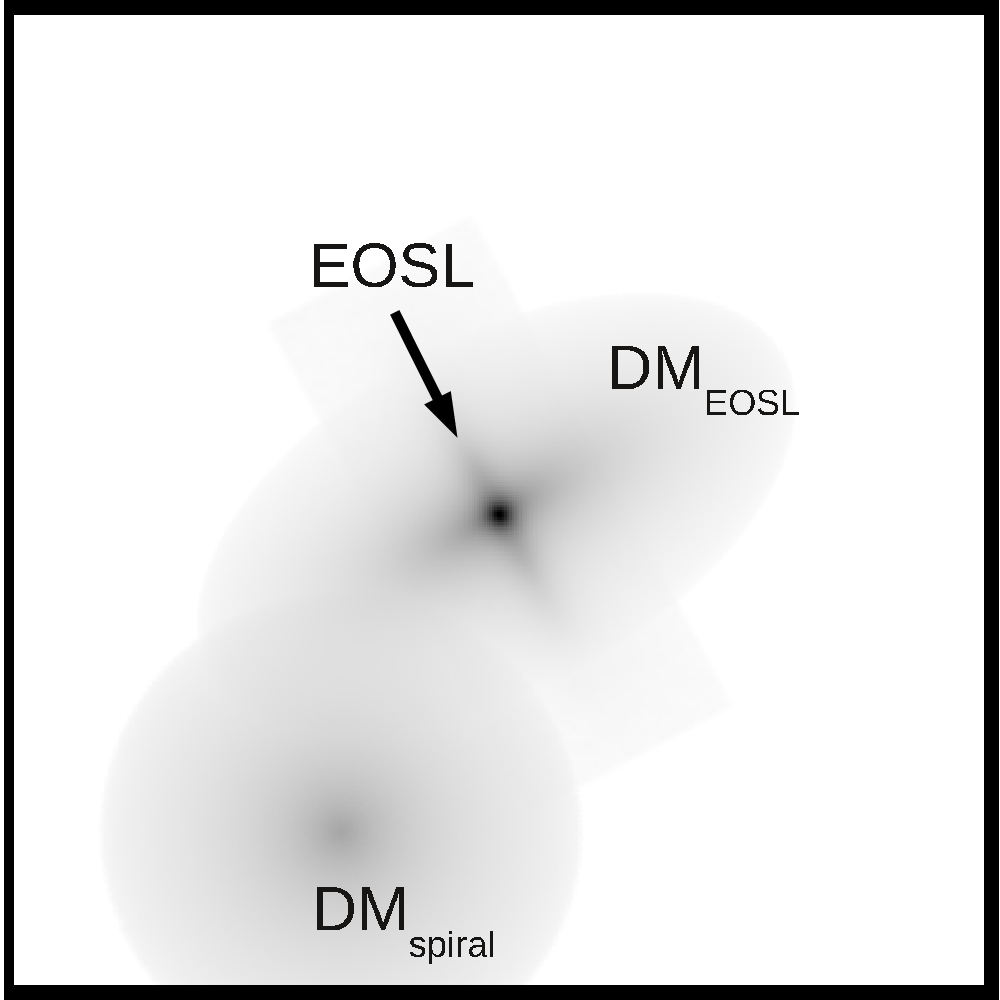}}
   \caption{Geometry of the lens plane. The three elements of the lens plane are marked. The EOSL DM corresponds to  
            model a in table 1 with $e=0.15$, and $\alpha=0^{\circ}$. } 
   \label{fig_geom_lens}  
\end{figure}

We should emphasize that the goal of this test is not to constrain the mass of the spiral galaxy but to test the robustness of our results and to identify changes in the 
best fitting model when some mass is allowed in the south sector of the EOSL so the particular choice of parameters for the spiral galaxy is not relevant for our purposes. 
Since our best fitting models are significantly determined by the slope of the lensed arc, structures that are farther than a few tens of kpc from the lensed arc have a very small 
impact on this data set and are mostly degenerate with respect to the properties of the background source, like the inclination angle $\theta$ and position $\Delta p$, as it will be discussed 
later. For clarity, we include in figure \ref{fig_geom_lens} the geometry of the lens plane including the 3 ingredients, the baryons in the EOSL, the EOSL DM for the case of $e=0.15$ and the 
spiral DM halo. 

\section{Results}\label{results}   

\begin{figure*}  
{\includegraphics[width=19.5cm]{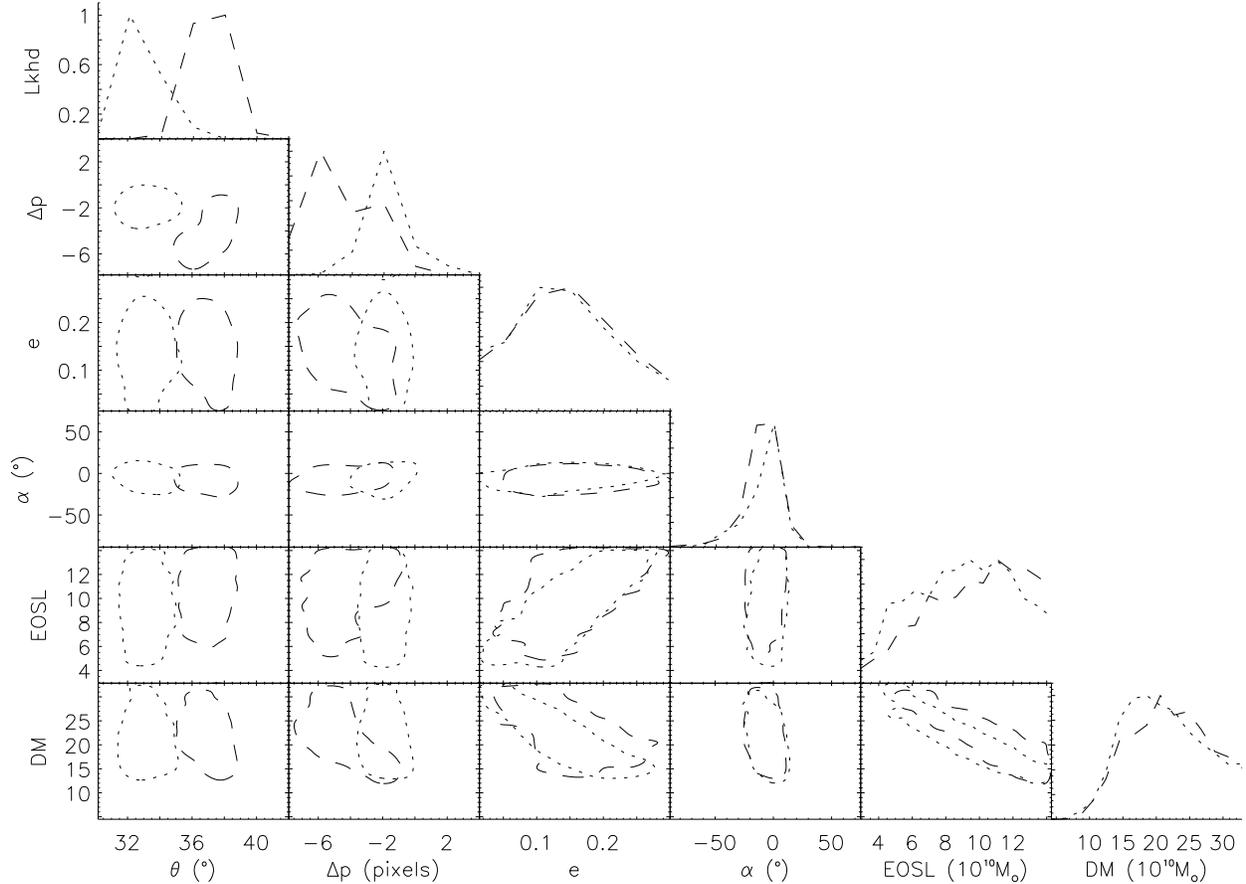}}
   \caption{Posterior probabilities for the parameters in our model. 
            The one dimensional plots correspond to the normalized (to 1) probability. The contours correspond to 68\% of the total 
            of the marginalized probability. 
            The dotted line probabilities are for the case when the mass of the spiral galaxy is fixed to $M_{spiral}=5\times 10^{10} M_{\odot}$ 
            and the dashed lines are for the case when  $M_{spiral}=22\times 10^{10} M_{\odot}$. Note how changes in the mass of the spiral basically 
            only affect the parameters of the background lensed source, $\theta$ and  $\Delta p$.  }
   \label{fig_Marg_prob_table}  
\end{figure*}

After computing the likelihoods for a grid of models (we compute two million models), there is a range of models that are consistent with the observed lensed arc.
In Figure \ref{fig_Fit_grid} we show a selection of particular models that are interesting to discuss in more detail. All models shown in this figure are listed in table 1. 
Model a (and e) is the best model that is found when the mass of the spiral is fixed to  $M_{spiral}=5\times 10^{10} M_{\odot}$ (or $M_{spiral}=22\times 10^{10} M_{\odot}$). 
Both the position of the arc and its derivative agree well with the observed arc and slope. For this model, the mass of the EOSL agrees better with the 
estimate of the stellar mass for the Salpeter IMF. The mass of the associated DM halo is relatively low compared with the baryonic component but we have to keep in mind 
that our reconstructed DM mass accounts mostly for the mass contained in the region that is sensitive to the lensing data. Our ability to cnstrain the mass distribution beyond the 
lensed arc is hence limited. The orientation of the DM halo is around 
$\alpha=0^{\circ}$ that corresponds to the long axis of the DM halo being perpendicular to the EOSL disc. The ellipticity takes reasonable values which are consistent 
with expectations. Forcing the EOSL galaxy to have a much smaller mass, like in model b (and f) results in an increase of the mass of the DM halo that compensates the reduction in 
baryonic mass and without affecting much the other parameters. 
A similar behaviour is observed when forcing the ellipticity to be very small (similar to the spherical halo) like in model c (and g). 
On the contrary, if we force the baryons to be the source of the lensing distortion like in case d (and h), the EOSL prefers a mass significantly higher than our stellar 
mass estimates but even in the best case, the fit is not satisfactory suggesting that this geometry is not able to reproduce the observations. Models d and h  
can be interpreted as a MOND-like model where no DM halo is needed to describe the observations. In a trully MOND model, the masses derived for the EOSL would be 
smaller than the ones in models d, and h and even probably consistent with the stellar mass but like in models d and h, a MOND model could not reproduce the 
observed arc as well as models that include a DM halo around the EOSL. This is an important result that directly challenges MOND-like models.  
 
Regarding the degeneracies between the different paramaters, we study them by marginalizing the likelihood over the otrher parameters. 
The results are shown in Figure \ref{fig_Marg_prob_table} for the two cases described above. 
The dotted line curves correspond to the case $M_{spiral}=5\times 10^{10} M_{\odot}$ and the dashed lines for the case   
$M_{spiral}=22\times 10^{10} M_{\odot}$. 
A few interesting conclusions can be made from these results. First, 
there is a degeneracy between $M_{EOSL}$ and $M_{DM}$ that can be approximated by the line $M_{EOSL}+M_{DM} \approx 3.0\times10^{11} M_{\odot}$.
This is not surprising. From this degeneracy, and to first order, the total mass of the galaxy can 
be well constrained. To second order, we see from the 1-dimensional probabilities that among the models with the same total mass, those that are more 
favoured are the ones with a baryon to DM ratio $\approx 0.5$  in agreement with recent findings based on microlensing, 
\citep{Jimenez2014}. It is important to note that this ratio accounts 
for the mass contained within a given radius and that it should not be regarded as the total mass of the galaxy.  
If the DM halo extends beyond the arc and with a symmetric distribution, the lensing distortion over the arc would be 
insensitive to the outer layers of mass. 
When comparing the derived lensing mass for the EOSL galaxy with the mass obtained photometrically (see beginning of section \ref{resultsII}),  
the best mass of the EOSL (from lensing) is in good agreement with the mass estimated photometrically, in particular with the case of the Salpeter mass function. 

Another very interesting conclusion is the fact that the data clearly favours models where the DM halo is aligned 
perpendicular to the disc of the galaxy ($\alpha=0^{\circ}$) as opposed to aligned with it ($\alpha=90^{\circ}$). 
We consistently find a prominent peak in the likelihood around values of $\alpha=0$. 
Such a clear preference for a perpendicular alignment between the baryonic disc and the DM challenges again models of modified gravity since a 
simple variation in $1/r^2$ could not account for this evident preference. Considering alternative laws to gravity 
would result only in changes of the gradient of the mass (with respect to the radius), 
but not on its relative distribution or geometry. 
The ellipticity is not constrained as well as the other halo parameters but there seems to be a preference for typical ellipticities $e \approx 0.15$ in good agreement with 
simulations. 

In general, the best parameters describing the EOSL (and its DM halo) show a weak dependency with the mass of the spiral. This is expected as 
any source of gravitational potential farther than the EOSL should contribute mostly as a gradient in the deflection angle at the position of the lensed arc. Such 
gradients are degenerate with the parameters describing the source parameters $\theta$ and $\Delta p$. 
This degeneracy is explicitly shown in the likelihoods of the orientation angle of 
the source $\theta$ and the relative  position $\Delta p$. The likelihoods of these parameters show the largest shifts when the mass of the spiral is changed. 
An increase in the spiral mass (or in the mass of S1) can be compensated by an increase in the tilt angle of the background source and/or a change in the relative position. 
Increasing the mass in the spiral brings the preferred value towards the centre of the EOSL (the centre of the EOSL corresponds to $\Delta p = -21$ pixels) 
in order to counter effect the increase in deflection angle from the spiral galaxy. 

\begin{figure}  
{\includegraphics[width=9cm]{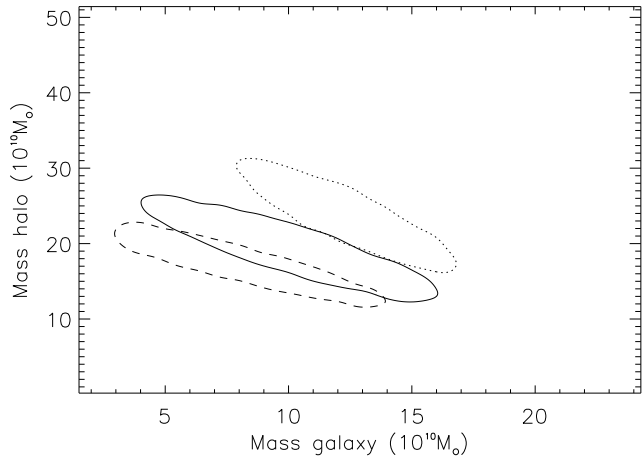}}
   \caption{Likelihood as a function of the masses of the galaxy and the halo for three different assumptions about the DM halo profile.  
            The three contours correspond to the 68\% confidence region. 
            The solid line corresponds to our standard DM halo with a small, concentration ($C=3$) and small virial radius ($R_{200} = 15$ kpc). 
            This is the case that was assumed for the main results in our paper. 
            The dashed line is for the case of a concentration three times larger ($c=9$) and the same virial radius ($R_{200} = 15$ kpc). The 
            dotted line  is for the case of small concentration ($C=3$) and larger virial radius ($R_{200} = 30$ kpc). }
   \label{fig_Lik1}  
\end{figure}

The derived constraints show also some dependency with the assumptions made about the 
parameters of the NFW profile used to simulate the DM halo. In Figure \ref{fig_Lik1} we show the constrains in the $M_{EOSL}-M_{DM}$ plane 
obtained when the concentration and virial radius are modified within reasonable limits. The remaining parameters are fixed to the values of model a listed in table 1. 
Changing the concentration parameter by a factor three results in a small shift along the degeneracy region. Increasing the virial radius by a factor 
two results in a higher total mass for the DM halo. However, this is mostly due to the fact that the DM halo is now larger. The mass 
enclosed in the region defined by the lensed arc remains more or less unchanged. In particular, comparing the best models obtained with the 15 kpc and 30 kpc radius 
and taking the mass of the larger halo that is enclosed in the footprint of the smaller halo, we find that both masses agree to within 3.5\%.    

In the results presented above we have assumed that the DM halo is prolate. We have checked that our results remain unchanged when we consider oblate models instead 
so we conclude that our test is insensitive to the differences between prolate and oblate models. 
Finally, in our results we have fixed the redshift of the lensed galaxy at $z_s=1$ (photo-z estimated with BPZ estimates the redshift for the background source 
between 0.7 and 1.2). This redshift is rather arbitrary however we should note that a different redshift would have an impact only on the masses which are degenerate with the 
redshift of the source. For example, for our particular cosmological model, assuming the background 
source is at $z_s=0.7$ would result in derived masses that are 39\% higher to compensate for the reduction in deflection angle. On the other hand, adopting 
$z_s=1.2$ would result in a decrease of the masses by 11\%.

\section{Discussion}
Perhaps the most important result is the strong preference for models that include DM. Also, the DM halo prefers to be aligned perpendicular to the plane 
of the galaxy ($\alpha=0$). N-body simulations predict that although this scenario is possible is not the most likely. 
\cite{Bailin2005} conclude that {\it the inner halo is aligned such that the halo minor axis aligns with 
the disk axis} in clear opposition with the findings in our work that suggest that it is the longest axis the one that 
is aligned with the disk axis. 
Also, in \cite{DeBuhr2012}, the authors use N-body simulations to study the relative orientation between the stellar 
discs and the DM haloes. They find that {\it regardless of the initial orientation of the disc, 
the inner parts of the haloes contract and change from prolate to oblate}. The authors explain also how 
this behaviour is found also when the initial condition is similar to the best model found in this work. 
According to the authors, when the major axis is aligned with the disc's normal, the length of the major axis 
quickly contracts and becomes the minor axis. More recently, using SDSS data, \cite{Loebman2014}, found that oblate models are also favoured for the halo 
around our Galaxy.

On the other hand, and although not necessarily connected with our results, a similar puzzling situation is observed with the distribution of 
satellites around galaxies. Recent studies based on the distribution of satellites around galaxies find that they concentrate in a thin disc in contradiction 
with expectations. In \cite{Ibata2013} the authors find a thin structure (with an inclination of $59^{\circ}$ from the galactic disc) 
around M31 which is at least 400 kpc in diameter, and extremely thin (perpendicular scatter of less than 14.1 kpc). 
Our own galaxy, exhibits a similar feature \cite{Kroupa2005}
with the satellites concentrating in the polar directions of the Milky Way (inclination of $88^{\circ}$ with respect to the Galactic plane, \cite{Metz2007}). 
Using N-body simulation, \cite{Bowden2013} study the likelihood of having such narrow discs of satellites and 
find that such configuration is only possible when the disc lies in the planes perpendicular to the long or short axis of a triaxial halo ($\alpha =0^{\circ}$ 
or $alpha=90^{\circ}$).  
In \cite{Zentner2005} and using N-body simulations, the authors find that sub-halos are distributed anisotropically and are preferentially located along 
the major axis of the triaxial mass distributions of their hosts. 
These numerical simulations, together with the observational evidence about the inclination of the satellite plane 
may support an scenario where the DM halo adopts a prolate distribution with its long axis 
perpendicular to the galactic disc, in a similar fashion to what our results seem to suggest. 

More sophisticated N-body simulations that focus for instance on the effects of tidal forces in the orientation 
of sub-haloes in clusters may be needed to answer this question. In \cite{Diego2014} the authors discuss how this cluster is undergoing a collision in a direction 
close to the line of sight and where the two cores are not merging face on but instead with a significant impact parameter. 
Simulations, and the fact that one of the X-ray peaks, and perhaps the peak of one the DM haloes, are displaced with respect to the dominant cD galaxies suggest 
the presence of strong tidal forces. Whether these tidal forces may have an impact in the orientation of the DM halo around the EOSL or even directly affect the stretch 
of the deflection field around the lensed arc is  an open question. 

An added advantage of the test presented in this work (based on EOSL) is that the results that can be derived are independent on whether gravity follows a 
$1/r^2$ law or not. Adopting a different law for gravity would result in different values for the masses of the haloes $M_{EOSL}$ and $M_{DM}$ but not on the geometry 
of the solution. 
Our findings clearly suggest the need of an additional halo of DM that does not trace the baryonic component. 

\section{Conclusions}
EOSL offer a unique opportunity to study the DM distribution around galaxies. 
When an EOSL is embeded in a massive galaxy cluster, the monopole lensing component from the cluster 
helps in magnifying the background galaxy, effectively stretching the background source behind the 
secondary lens galaxy. The stretch provided by the galaxy cluster helps constraining the dark matter by reducing 
some of the degeneracies inherent to lensing reconstruction but also by producing a straight magnified background 
source that can be used to sample the lensing potential of the secondary lens to larger distances. 
We apply the EOSL test to a galaxy found in the MACSJ0416.1-2403 galaxy cluster. We take advantage of our previous 
work where the galaxy cluster lens model is determined with accuracy to constrain the properties of the arc that 
is being lensed by the EOSL. We explore the space of solutions with a model containing 6 parameters (and an additional halo for the nearby 
spiral galaxy). After marginalizing over the space parameter  we are able to constrain the total mass of the galaxy although some 
degeneracies still persist between the baryonic and DM masses. However, regarding the spatial distribution of 
the DM, the marginalized probability shows a strong preference for prolate (or oblate) models that contain the bulk of the mass  
in a DM halo that is perpendicular to the plane of the visible galaxy. This scenario, although  
consistent with some simulations, would contradict also other simulations, that predict that most of the 
times the inner part of the DM halo aligns in a direction that is in line with the visible galaxy.  
The existence and geometry of the elongated DM halo, perpendicular to the galaxy, goes against the hypothesis of modified gravity theories. In 
these models, models aligned with the visible galaxy would be favoured if the baryonic component is responsible for the gravitational potential. 
A simple change in the Newtonian potential can not eliminate the need for a dark component and would result only in different mass estimates but still 
with a preference for a dark component that aligns perpendicularly with the emitting light.

\section{Acknowledgments}  
This work is based on observations made with the NASA/ESA {\it Hubble Space Telescope} and operated by the Association of Universities for Research in Astronomy, Inc. under NASA contract NAS 5-2655. T.J.B. thanks the University of Hong Kong for generous hospitality. J.M.D acknowledges support of the consolider project CAD2010-00064 and AYA2012-39475-C02-01 funded by the Ministerio de Economia y Competitividad. J.M.D also acknowledges the hospitality of the Department of Physics and Astronomy at Univeristy of Penn during part of this research. 
  
\label{lastpage}
\bibliographystyle{mn2e}
\bibliography{MyBiblio} 

\begin{thebibliography}{17}
\expandafter\ifx\csname natexlab\endcsname\relax\def\natexlab#1{#1}\fi

\bibitem[{{Bailin} {et~al}\mbox{.}(2005){Bailin}, {Kawata}, {Gibson},
  {Steinmetz}, {Navarro}, {Brook}, {Gill}, {Ibata}, {Knebe}, {Lewis}, \&
  {Okamoto}}]{Bailin2005}
{Bailin} J. {et~al.}, 2005, \apjl, 627, L17

\bibitem[{{Bowden}, {Evans} \& {Belokurov}(2013){Bowden}, {Evans}, \&
  {Belokurov}}]{Bowden2013}
{Bowden} A., {Evans} N.~W., {Belokurov} V., 2013, \mnras, 435, 928

\bibitem[{{Clifton} {et~al}\mbox{.}(2012){Clifton}, {Ferreira}, {Padilla}, \&
  {Skordis}}]{Clifton2012}
{Clifton} T., {Ferreira} P.~G., {Padilla} A., {Skordis} C., 2012, Phys. Rep,
  513, 1

\bibitem[{{DeBuhr}, {Ma} \& {White}(2012){DeBuhr}, {Ma}, \&
  {White}}]{DeBuhr2012}
{DeBuhr} J., {Ma} C.-P., {White} S.~D.~M., 2012, \mnras, 426, 983

\bibitem[{{Despali}, {Giocoli} \& {Tormen}(2014){Despali}, {Giocoli}, \&
  {Tormen}}]{Despali2014}
{Despali} G., {Giocoli} C., {Tormen} G., 2014, \mnras, 443, 3208

\bibitem[{{Diego} {et~al}\mbox{.}(2014){Diego}, {Broadhurst}, {Molnar}, {Lam},
  \& {Lim}}]{Diego2014}
{Diego} J.~M., {Broadhurst} T., {Molnar} S.~M., {Lam} D., {Lim} J., 2014, ArXiv
  e-prints 1406.1217

\bibitem[{{Ibata} {et~al}\mbox{.}(2013){Ibata}, {Lewis}, {Conn}, {Irwin},
  {McConnachie}, {Chapman}, {Collins}, {Fardal}, {Ferguson}, {Ibata}, {Mackey},
  {Martin}, {Navarro}, {Rich}, {Valls-Gabaud}, \& {Widrow}}]{Ibata2013}
{Ibata} R.~A. {et~al.}, 2013, \nat, 493, 62

\bibitem[{{Jauzac} {et~al}\mbox{.}(2014){Jauzac}, {Jullo}, {Eckert}, {Ebeling},
  {Richard}, {Limousin}, {Atek}, {Kneib}, {Cl{\'e}ment}, {Egami}, {Harvey},
  {Knowles}, {Massey}, {Natarajan}, \& {Rexroth}}]{Jauzac2014}
{Jauzac} M. {et~al.}, 2014, ArXiv e-prints 1406.3011

\bibitem[{{Jim{\'e}nez-Vicente} {et~al}\mbox{.}(2014){Jim{\'e}nez-Vicente},
  {Mediavilla}, {Kochanek}, \& {Mu{\~n}oz}}]{Jimenez2014}
{Jim{\'e}nez-Vicente} J., {Mediavilla} E., {Kochanek} C.~S., {Mu{\~n}oz} J.~A.,
  2014, ArXiv e-prints

\bibitem[{{Jing} \& {Suto}(2002)}]{Jing2002}
{Jing} Y.~P., {Suto} Y., 2002, \apj, 574, 538

\bibitem[{{Khoury}(2014)}]{Khoury2014}
{Khoury} J., 2014, ArXiv e-prints

\bibitem[{{Kroupa}, {Theis} \& {Boily}(2005){Kroupa}, {Theis}, \&
  {Boily}}]{Kroupa2005}
{Kroupa} P., {Theis} C., {Boily} C.~M., 2005, \aap, 431, 517

\bibitem[{{Loebman} {et~al}\mbox{.}(2014){Loebman}, {Ivezic}, {Quinn}, {Bovy},
  {Christensen}, {Juric}, {Roskar}, {Brooks}, \& {Governato}}]{Loebman2014}
{Loebman} S.~R. {et~al.}, 2014, ArXiv e-prints

\bibitem[{{Metz}, {Kroupa} \& {Jerjen}(2007){Metz}, {Kroupa}, \&
  {Jerjen}}]{Metz2007}
{Metz} M., {Kroupa} P., {Jerjen} H., 2007, \mnras, 374, 1125

\bibitem[{{Milgrom}(1983)}]{Milgrom1983}
{Milgrom} M., 1983, \apj, 270, 365

\bibitem[{{Sanders} \& {McGaugh}(2002)}]{Sanders2002}
{Sanders} R.~H., {McGaugh} S.~S., 2002, \araa, 40, 263

\bibitem[{{Zentner} {et~al}\mbox{.}(2005){Zentner}, {Kravtsov}, {Gnedin}, \&
  {Klypin}}]{Zentner2005}
{Zentner} A.~R., {Kravtsov} A.~V., {Gnedin} O.~Y., {Klypin} A.~A., 2005, \apj,
  629, 219

\end{thebibliography}

 
\end{document}